\documentclass{iopart}

\usepackage{graphicx}
\usepackage[squaren]{SIunits}

\begin{document}

\title{Single Defect Centers in Diamond Nanocrystals as Quantum Probes for Plasmonic Nanostructures}

\author{Andreas W. Schell,$^\text{a}$* G\"{u}nter Kewes,$^\text{a}$ Tobias Hanke,$^\text{b}$ Alfred Leitenstorfer,$^\text{b}$ Rudolf Bratschitsch,$^\text{b}$ Oliver Benson$^\text{a}$ and Thomas Aichele$^\text{a}$}
\address{$^\text{a}$  Humboldt-Universit\"{a}t zu Berlin, Institute of Physics,\\
Newtonstra{\ss}e~15, D-12489 Berlin, Germany \\ $^\text{b}$ University of Konstanz, Department of Physics and Center for Applied Photonics, \\ Universit\"{a}tsstra{\ss}e 10, D-78464 Konstanz, Germany
*andreas.schell@physik.hu-berlin.de}

\begin{abstract}
We present two applications of a single nitrogen vacancy center in a nanodiamond as quantum probe for plasmonic
nanostructures. Coupling to the nanostructures is achieved in a highly controlled manner by picking up a pre-characterized nanocrystal
with an atomic force microscope and placing it at the desired position. Local launching of single excitations
into a nanowire with a spatial control of few nanometers is demonstrated. Further, a two dimensional map of the
electromagnetic environment of a plasmonic bowtie antenna was derived, resembling an ultimate limit of
fluorescence lifetime nanoscopy.
\end{abstract}

Defect centers in diamond have proven to be excellent quantum light sources. They are active at room
temperature, do not photobleach and have long spin decoherence times even at room temperature
\cite{Kennedy2003}. In particular the nitrogen vacancy (NV) center \cite{Jelezko2006,Brouri2000} has been widely studied
to realize single photon emitters \cite{Kurtsiefer2000,Schroder2010} or nanophotonic elements
\cite{Wolters2010}, to implement quantum information processing \cite{Ladd2010}, and to optically 
detect magnetic fields with nanometer spatial resolution \cite{2008Natur.455..648B,Maze2008}.
The latter publications have introduced NV centers as optical nanoprobes. NV centers in diamond nanocrystals are
particularly useful for this application due to their small size of less than a few 10 nanometers \cite{Sonnefraud2008}.
STED microscopy with NV centers for example provides an optical resolution of 6 nm \cite{Rittweger2009}. 
The role of an optical nanoprobe can be twofold. First, it can create a well defined optical
excitation which is then launched into a nearby system under investigation \cite{Cuche2010}. The generation of a single photon
from an NV center can thus be regarded as an ultimate quantum limit of a pump pulse in a pump-probe experiment.
Second, the probe itself can respond to a change of its local environment which can then be monitored
optically. Key requirements for a local quantum probe are stability, since only a weak signal is created, and
the ability to position the probe with excellent spatial precision. In this paper we demonstrate how NV centers
in single nanodiamonds can be utilized as quantum probes to study plasmonic nanostructures.

Surface plasmons polaritons (SPPs) \cite{Barnes2003} concentrate electromagnetic fields in volumes much
smaller than the optical wavelength. This can be exploited to enhance light-matter interaction and to modify the
emission characteristics of light emitters. However, due to the strong confinement sub-wavelength probing is
mandatory if such structures shall be characterized satisfactorily. In a first experiment, we utilize the
nanodiamond quantum probe to study propagation of SPPs along metal nanowires \cite{Ditlbacher2005}. Propagation loss, as well as
input/output efficiency are crucial parameters for applications of SPP waveguides in novel miniaturized photonic
components \cite{Li2010,Knight2007}. In the following we study SPPs on the level of single quanta. In previous
experiments, excitation of single surface plasmon states by a single NV center and guiding along a metal
nanowire was demonstrated using structures which were randomly assembled by deposition of metal nanowires and
quantum emitters on a sample surface \cite{Kolesov2009,Akimov2007,Fedutik2007,Wei2009}. Therefore, on-demand positioning and
change of a once assembled configuration was not possible.

\begin{figure}[h]
  \centering
  \includegraphics[width=0.9\textwidth]{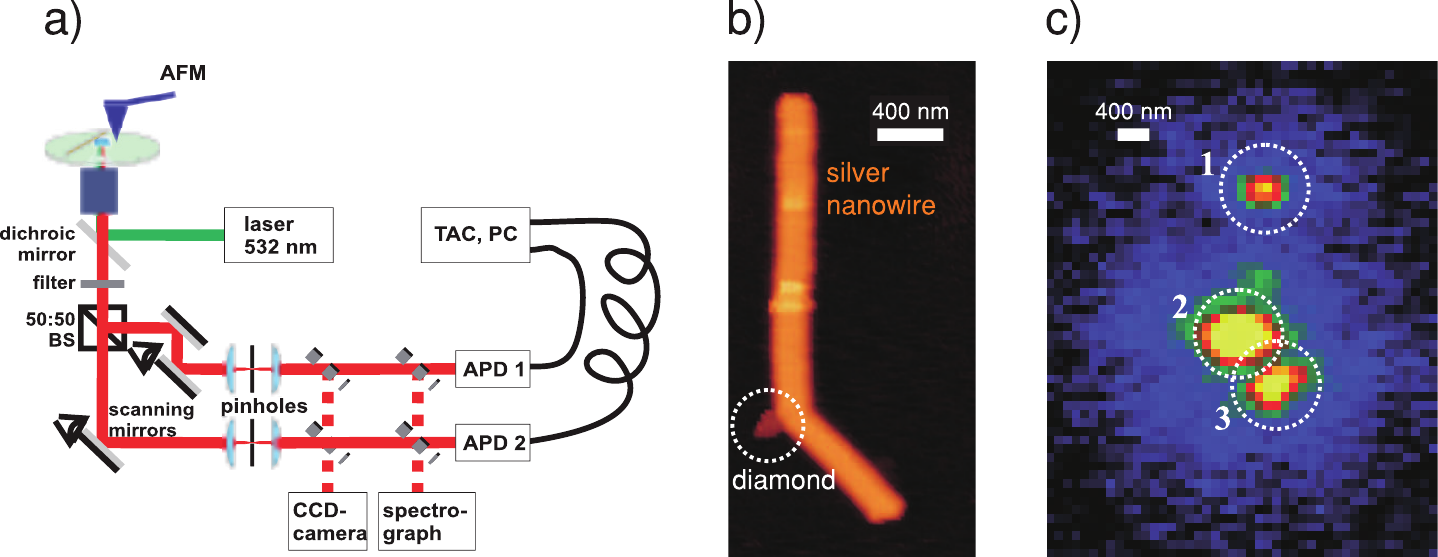}
  \caption{Coupling of a diamond nanocrystal with a single NV center to a silver nanowire and excitation of
  single surface plasmon polaritons. (a) Scheme of the experimental setup. (b) AFM image of the diamond nanocrystal
  (indicated by the dashed circle)
  positioned at the side of a bent silver nanowire. (c) Microscope photoluminescence image of the  same configuration.
  Positions~1 and 3 indicate the ends of the nanowire, while position~2 marks the location of the diamond nanocrystal.
  }
  \label{fig:nanowire}
\end{figure}

In our experiments, we use a setup consisting of a confocal microscope combined with an atomic force microscope
(AFM) (see Fig. \ref{fig:nanowire}(a)). This combination allows both optical
detection of photons and nanomanipulation of the nanodiamond probes \cite{Barth2009,Schietinger2009}. 
Photon correlation measurements were performed with a Hanbury-Brown and Twiss (HBT) configuration of two
avalanche photo diodes (APDs, quantum efficiency at $\unit{700}{\nano\meter}$ approximately $30~\%$). A diamond nanocrystal
containing a single NV defect center and thus resembling a quantum probe was first optically characterized on a
coverslip and then individually picked up by the AFM tip. Then, the nanocrystal was transferred to a coverslip
with chemically synthesized silver nanowires and placed on-demand near a previously selected wire which served
as SPP waveguide. 
With this technique one can be sure that there is exactly one diamond containing exactly one single NV defect center on the whole sample.
So there is no possibility of accidentally measuring photons coming from a diamond containing more than one defect center. 
Nanomanipulation with the AFM tip then allowed positioning of the nanocrystal and launching of
a single excitation at arbitrary positions along the wire.

\begin{figure}[h]
  \centering
  \includegraphics[width=0.9\textwidth]{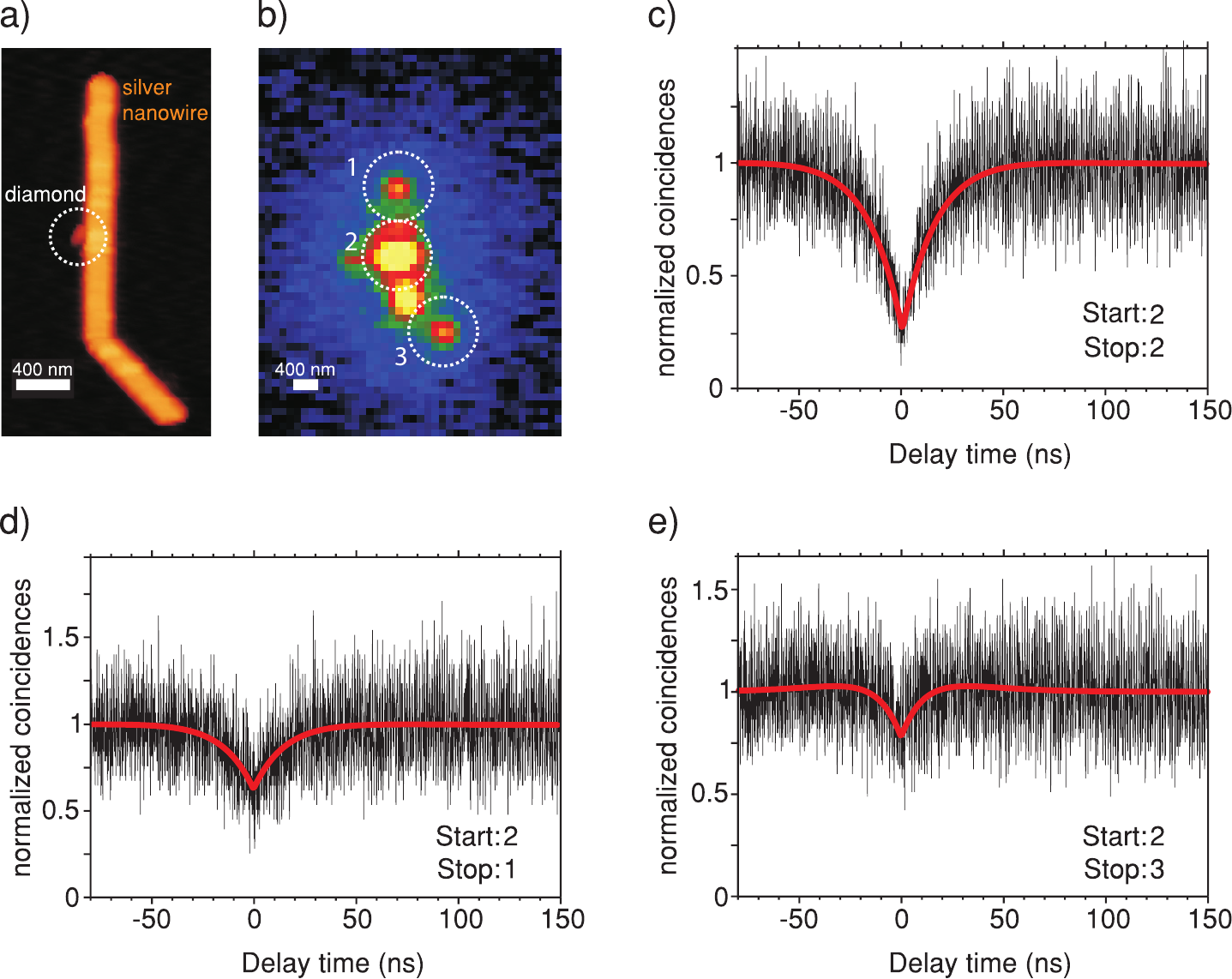}
  \caption{A nanodiamond probe launching single plasmonic excitations. (a,b) AFM and fluorescence image
  showing a silver nanowire and a nanodiamond with a single NV center. The bright spot between position~2 and 3 in (b)
  emerges from the bend of the nanowire. (c) Autocorrelation of the photons from the diamond measured at point 2.
  (d,e) Cross-correlations between the photons emitted from position 2 and from positions 1 and 3, respectively.
  The anti-bunching dip in the curves clearly reveals the non-classical properties.}
  \label{fig:nanowirecorrelation}
\end{figure}

Fig. \ref{fig:nanowire}(b) shows an AFM image of the nanowire, which has a diameter of \unit{\approx
80}{\nano\meter}. A sharp bend separates the nanowire into two arms of \unit{1.9}{\micro\meter} and
\unit{0.7}{\micro\meter} length. Under continuous wave laser excitation of the NV center (at a wavelenght of
\unit{532}{\nano\meter}), photoluminescence directly from the diamond nanocrystal (position~2 in Fig. \ref{fig:nanowire}(c)) 
as well as light emerging from the bend and from the ends of the nanowire
(positions~1 and 3 in Fig. \ref{fig:nanowire}(c)) are visible. Since there is a strong fluorescence background
emerging from the bend of the nanowire while exciting the diamond, for further measurements the diamond was
placed at another position, so that the nanowire bend is no longer in the excitation spot (position 2 in
 Fig. \ref{fig:nanowirecorrelation}(c)). Already this repositioning of the nanocrystal shows the advantage of
nanomanipulation. Accidental inconvenient configurations can be corrected and experiments can be repeated under
otherwise unchanged conditions. At the same time, the ability to reposition is crucial in more complex
structures, since slight changes of the position of an emitter with respect to a plasmonic structure may already
modify its emission as well as the structures plasmonic properties significantly. 
Despite the possibility to perform near-field simulations, under experimental conditions
an a priori prediction of an optimum position for an emitter or nanoprobe
is often impossible \cite{Chang2006}.

With the new position of the diamond nanocrystal there are now four fluorescent spots observable on the
nanowire. One (position 2 in Fig. \ref{fig:nanowirecorrelation}(b)) corresponds to the emission of fluorescence
directly from the nanodiamond. An autocorrelation measurement on this spot shows a pronounced anti-bunching
behavior (Fig. \ref{fig:nanowirecorrelation}(c)), confirming the single photon character of emission from a
single NV center in the nanodiamond. The other spots now correspond to three output ports for single
excitations launched into the wire via the nanodiamond quantum probe. In order to prove the quantum nature of
the excitations we performed cross-correlation measurements between the light directly emitted from the
nanodiamond (position 2 in Fig. \ref{fig:nanowirecorrelation}(b)) and the light emitted from both ends of the
nanowire (positions 1 and 3, respectively, in Fig. \ref{fig:nanowirecorrelation}(b)). Again, a clear
anti-bunching was observed proving that indeed light from the quantum probe was converted into single plasmonic
excitations which propagated to the output ports. The nanowire thus represents a plasmonic beamsplitter with
three output ports, a key building block for quantum plasmonic elements. When generating plasmons from the
excitation of a single quantum emitter anti-bunching after reconversion to photons is expected. However, we
would like to point out that the preservation of a pronounced anti-bunching dip is an excellent indicator to
quantify the contribution of unwanted stray light and background fluorescence which may occur in complex plasmonic structures. The single
nanodiamond is also a perfect test probe for this purpose.

In another experiment with optical antennas we exploit the second feature of  the nanodiamond probe, i.e. 
its ability to map the local electromagnetic environment of a plasmonic nanostructure.

A plasmonic antenna is highly desired in experiments on the level of single quantum emitters as it excites the
quantum emitter efficiently, extracts many photons out of the quantum emitter, and spatially directs the emitted
photons towards a specific direction \cite{Muhlschlegel2005,Muskens2007,Esteban2010}. 
We investigated gold bowtie antennas fabricated with electron-beam lithography on a glass substrate 
with our nanodiamond probe. In contrast to a configuration where the antenna is decorated with several
emitters in a purely random manner \cite{Kinkhabwala2009}, we utilize again AFM nanomanipulation. In this way we
control the position of the nanodiamond with nanometer precision, i.e. we investigate a large number of
different emitter/antenna configurations, using exactly the same constituents \cite{MerleinJorg2008}. A mapping of the electromagnetic
environment via observing the optical properties of the nanoprobe is thus possible.

In the experiment we used the setup shown in Fig. \ref{fig:nanowire}(a), but with another excitation laser
operating at \unit{10}{\mega\hertz} pulsed excitation at \unit{532}{\nano\meter} wavelength. All measurements
(except the power dependence) were performed at an power of \unit{40}{\micro\watt} at the back of the slightly
overfilled NA=1.35 oil immersion objective.

\begin{figure}[htbp]
  \centering
  \includegraphics[width=0.9\textwidth]{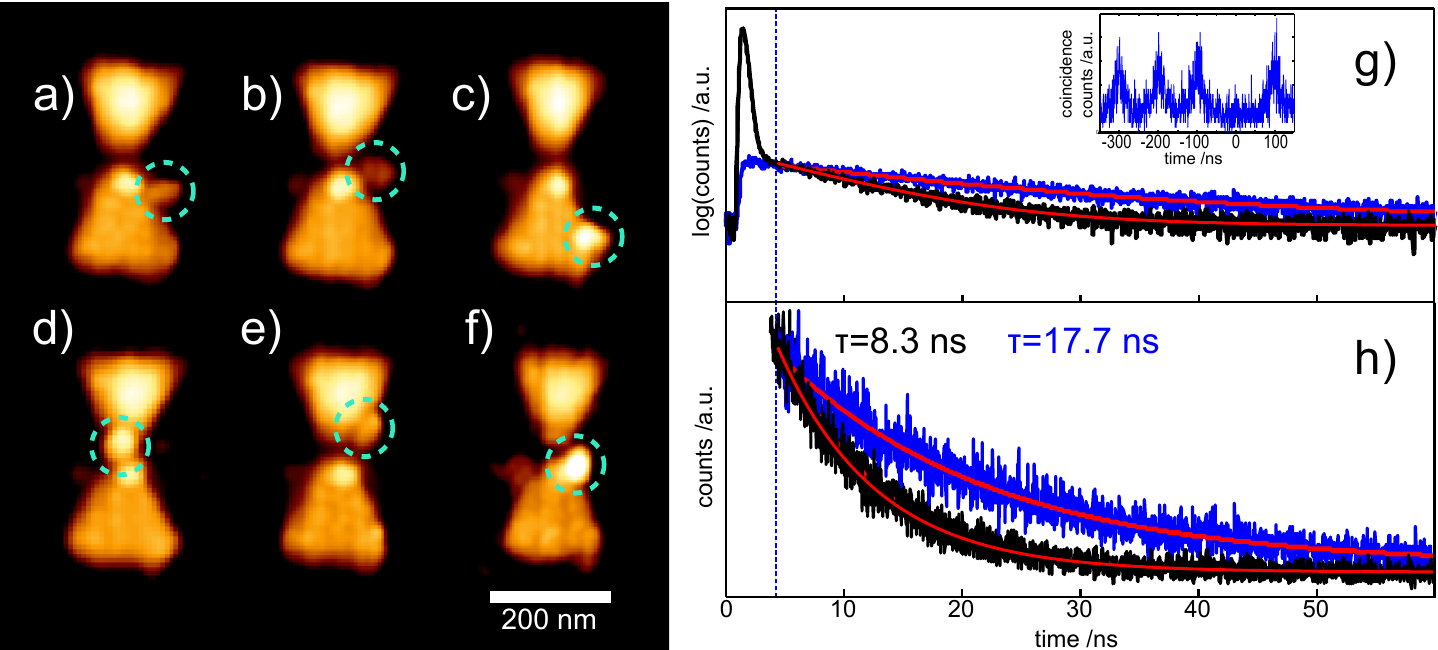}
  \caption{(a-f) Aligning a nanodiamond probe. A single nanodiamond was moved with the AFM to different positions
  with respect to a bowtie antenna. Small changes in the measured topography are due to changes of the AFM tip's shape
  during the manipulation process. Circles highlight the positions of the nanodiamond.
  (g,h) Fluorescence lifetime histogram of a uncoupled nanodiamond (blue) and the nanodiamond coupled to a bowtie antenna 
  (black). The initial peak stems from short-lived fluorescence from the gold of the bowtie antenna, 
  which our detectors could not resolve. Only counts occurring $\unit{3}{\nano\second}$ after the emission peak 
  (indicated by the blue vertical line) are used to fit (red curves) the fluorescence decay from the NV center in the nanodiamond.
  Inset in (g) shows the coincidences for the uncoupled diamond measured with the HBT setup.}
  \label{fig:diamondpositions}
\end{figure}

\begin{figure}[htbp]
  \centering
  \includegraphics[width=0.9\textwidth]{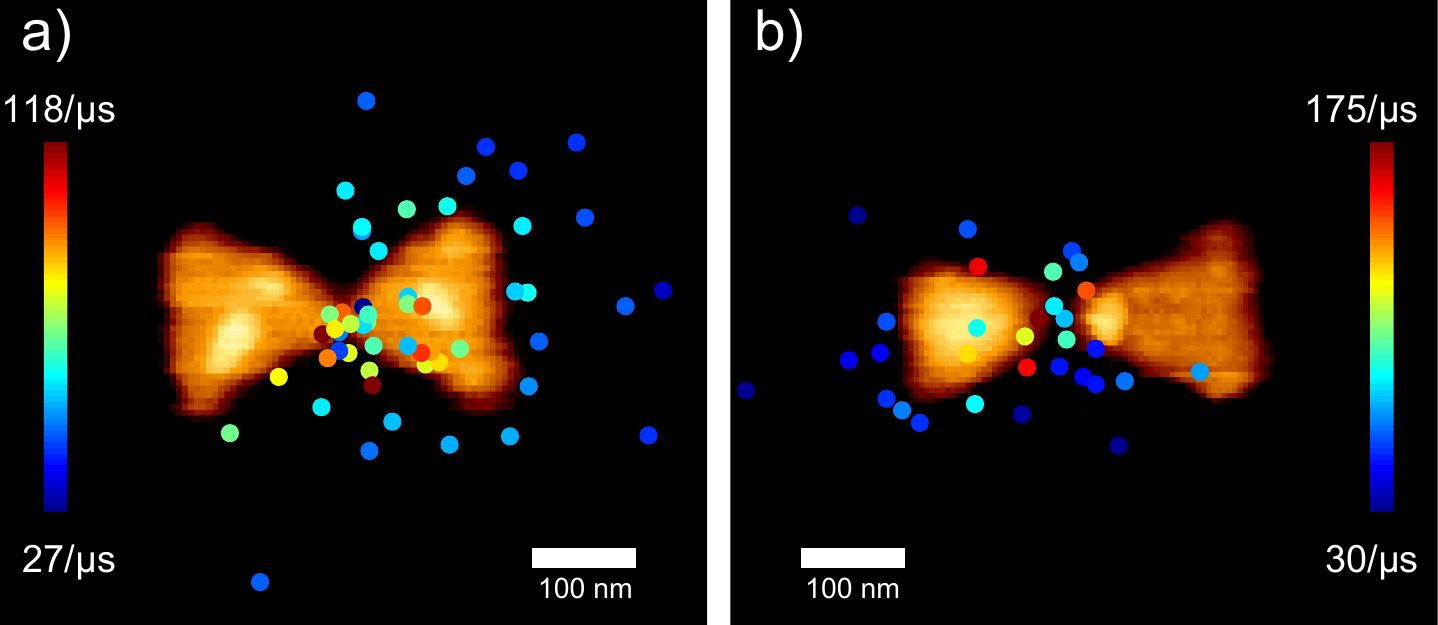}
  \caption{(a,b) Maps showing the decay rate of the excited state of the NV center in the nanodiamond probe for
  different positions with an underlay of the corresponding AFM image of the antennas.
  In (a) a diamond of approx.~$\unit{60}{\nano\meter}$ height and an oval shape was used.
  In (b) the diamond height was about $\unit{15}{\nano\meter}$.}
  \label{fig:decaymaps}
\end{figure}

There are different processes when coupling an emitter to a plasmonic antenna. The excitation as well as the
radiative rate are enhanced, but additional non-radiative decay channels may open up \cite{Farahani2005}. Also
the modified spatial emission pattern may change the number of the detected photons on a detector of finite
solid angle \cite{Curto2010}. In our spatial mapping we determined the lifetime of the NV center's excited
state and the total enhancement of the photon emission rate. 
By repeatedly measuring the lifetime and changing the nanodiamond's position with the AFM (see
Fig. \ref{fig:diamondpositions}(a-f)) we obtained the lifetime maps depicted in Fig. \ref{fig:decaymaps} (a,b).
The lifetimes were measured via time correlated single photon counting. In order to suppress short-lived
emission from the gold, count events within the first few nanoseconds were not taken into account for the
fitting to theoretical decay curves (see Fig. \ref{fig:diamondpositions} (g,h)).

The antennas were fabricated to have a gap of \unit{10}{\nano\meter} and consist of two isoscele triangles with
an altitude that equals the short side of \unit{140}{\nano\meter} (Fig. \ref{fig:decaymaps}(a)) and
\unit{150}{\nano\meter} (Fig. \ref{fig:decaymaps}(b)). This geometries give a fundamental mode in the infra-red. 
In Fig. \ref{fig:decaymaps}(a) a diamond with a height of 
approximately \unit{60}{\nano\meter} and an oval shape was used, which was too large to fit in the antenna gap.
The exact position of the NV center in this nanodiamond has therefore also an a priori uncertainty of
approx.~\unit{60}{\nano\meter}. For this reason in the measurements depicted in Fig. \ref{fig:decaymaps}(b) we
used a nanodiamond of \unit{15}{\nano\meter} height to reduce these problems. The two-dimensional maps in 
Fig. \ref{fig:decaymaps} resemble an ultimate limit of fluorescence lifetime nanoscopy, having only one quantum 
emitter present.

Beyond the mapping of the electromagnetic environment of the antenna our data provides information about the
actual enhancement of the fluorescence rate of the NV center in the nanodiamond. For a quantitative analysis, we
compared the uncoupled diamond to the situation where the nanodiamond is coupled to the antenna in an optimum
position. We placed the nanodiamond in the antenna gap, where also an excellent alignment of the diamond with
respect to the excitation laser spot is possible. The uncoupled
diamond has a lifetime of $\tau_{u} = \unit{17.7}{\nano\second}$, which is reduced to $\tau_{c} =
\unit{8.3}{\nano\second}$ when the diamond is in the antenna gap. Also the photon emission rate from the diamond is
changed from $\text{R}_{u} = \unit{2.5}{\kilo\hertz}$ to $\text{R}_{c} =\unit{2.2}{\kilo\hertz}$. The reduction
is due to additional loss channels which open up close to the metal surface. However, care has to be taken in the
analysis since we perform pulsed excitation at a fixed excitation rate and a single NV center only
can get excited once per cycle, if the duration of the excitation pulse is small compared to the NV center's lifetime. 
Due to coupling to the antenna both radiative and non-radiative decay is
enhanced, i.e. the lifetime of the excited state is significantly shortened. This means that although the
probability of generating a photon after excitation is reduced, photons can be provided in principle at a much
higher rate. 

The key number is the rate of photon emission under continuous saturated excitation, which could be calculated from the 
emission's power dependence under pulsed excitation. Since we observed melting of the gold nanoantennas at an excitation 
power of ca. $\unit{50}{\micro\watt}$, which is clearly below the saturation intensity, it was not possible to determine 
the enhancement factor with adequate accuracy.

In summary, we have shown that a single NV center in a nanodiamond can act as stable quantum probe working at
room temperature. Together with a controlled AFM manipulation ultimate limits of pump-probe experiments can be
performed, where the pump consists of only a single excitation launched with nanometer spatial precision into
the structure under investigation, in this case a plasmonic beamsplitter. At the same time two-dimensional maps
of plasmonic antenna structures can be derived. These maps provide insight into the near-field properties of
antenna structures allowing optimization of the designs \cite{Wissert2009}. 
The method of
repeatedly repositioning the quantum emitter and mapping the lifetime can in principle be extended to almost any
system of interest, if they are accessible with an AFM \cite{Wolters2010,Schroder2010}.

We thank Ulrike Woggon for providing the silver nanowires and the DFG project IQuOSuPla for financial support.
\\
\\
\\

\end{document}